\documentclass{ws-p8-50x6-00}

\begin{document}


\title{Large $N_{c}$ means $N_{c}$ = 3}

\author{M.D. Scadron}
\address{Physics Department, University of Arizona, Tucson, AZ 85721, USA}


\maketitle

\section{Introduction}

It is well known\cite{r1} that the interacting part of the SU(2)
linear $\sigma$ model (L$\sigma$M) Lagrangian density (after the
spontaneous symmetry breaking shift) is:
\begin{equation} \label{e11}
{\cal L}^{int}_{L\sigma M}= g \overline{\psi}(\sigma + i \gamma_{5}
\vec{\tau} \cdot \vec{\pi})\psi \ +\ g'\sigma(\sigma^{2}+
\vec{\pi}^{2}) -\ \lambda(\sigma^{2} + \vec{\pi}^{2})^{2}/4,
\end{equation}
where the couplings $g$, $g'$, $\lambda$ obey the chiral L$\sigma$M
relations
\begin{equation} \label{e12}
g = m_{q}/f_{\pi}, \ \ \ \ \ \ \ \ \ \ \ g' = m^{2}_{\sigma}/ 2
f_{\pi} = \lambda f_{\pi}.
\end{equation}
Here $f_{\pi} \approx 93$ MeV, and we take the fermion spinors as
quark fields.  In the chiral limit (CL) a once-subtracted dispersion
relation combined with unitarity predicts\cite{r2}
\begin{equation} \label{e13}
1 \ -\ \frac{f^{CL}_{\pi}}{f_{\pi}} \ = \ \frac{ m^{2}_{\pi} }{
8\pi^{2}f^{2}_{\pi} }\ \approx \ 0.03,
\end{equation}
so the observed pion decay constant $f_{\pi} \approx 93$ MeV requires
$f_{\pi}^{CL} \approx$ 90 MeV.  While the chiral relations (\ref{e12})
already hold at tree level\cite{r1}, they also remain valid at
one-loop level along with two additional L$\sigma$M
relations\cite{r3}:
\begin{equation} \label{e14}
m_{\sigma} = 2 m_{q}, \ \ \ \ \ \ \ \ \ \ \ \ \ \ \ g = g_{\pi q q }=
g_{\sigma qq} = 2\pi/ \sqrt{N_{c}}.
\end{equation}
The former relation in (\ref{e14}) is the famous NJL relation\cite{r4}
and the latter equation in (\ref{e14}) for $N_{c}=3$ gives $g = 2\pi
/\sqrt{3}=3.6276$, so the quark-level Goldberger-Treiman relation
(GTR) in (\ref{e12}) predicts
\begin{equation} \label{e15a}
m_{q}^{CL}\ =\ f_{\pi}^{CL}g\ \approx \ (90 \ {\rm MeV})2\pi/\sqrt{3}\
\approx \ 325 \ {\rm MeV},
\end{equation}
reasonably near the expected nonstrange constituent quark mass $m_{q}
\approx M_{N}/3 \approx 313$ MeV.  Also, the 3 constituent quarks
in a nucleon (uud or ddu) then find the predicted $\pi$qq coupling $g
\ = \ 2\pi/\sqrt{3}\ \approx \ 3.63$ to be nearby
\begin{equation} \label{e15b}
 g\  =\  g_{\pi N N} / 3 g_{A} \  \approx \ 3.47,
\end{equation}
for the measured $g_{\pi NN}\approx 13.2$, $g_{A} \approx 1.267$ and
near the GTR value $g = m_{q}/ f_{\pi}^{CL} \approx 313 \ {\rm MeV}/ \
90 \ {\rm MeV} \ \approx \ 3.48$.  Moreover, $g\ = \ 2\pi/\sqrt{3}$ in
Eqs.~(\ref{e14}), (\ref{e15a}), (\ref{e15b}) also follows from the Z =
0 compositeness condition\cite{r5,r6}.

\section{Color Number $N_{c}=3$ in the SU(2) L$\sigma$M}

It is well understood that the quark model must have an additional
quantum number (color) in order to describe qqq decuplet as well as
qqq octet baryons.  The phenomenological determination of $N_{c}$ is
from the axial anomaly\cite{r7} (plus PCAC) or from the PVV L$\sigma$M
u,d quark loop\cite{r8}.  Then the $\pi^{0} \rightarrow 2\gamma$ decay
rate of 7.7 eV implies that the amplitude magnitude satisfies
\begin{equation} \label{e21}
|M_{ \pi^{0} \gamma \gamma }| \ = \ \frac{N_{c}}{3} \frac{\alpha}{\pi
f_{\pi}} \ = \ (0.025 \pm 0.001) \ {\rm GeV}^{-1}.
\end{equation}
But $\alpha/\pi f_{\pi}\ \approx 0.0025 \ {\rm GeV}^{-1}$ for $f_{\pi}
\approx 93 \ {\rm MeV}$, so that Eq.~(\ref{e21}) suggests $N_{c}/3 =1$ or
$N_{c}=3$.

The SU(2) L$\sigma$M theoretical value of $N_{c}$ stems from the
Ben~Lee null tadpole sum\cite{r9} (due to u,d quark loop, $\vec{\pi}$
loop and $\sigma$ loop tadpole graphs linked to the scalar $\sigma$
meson as shown in Fig.~1):
\begin{equation} \label{e22}
<\sigma> \ =\ -8N_{c}m_{q}g \int\frac{d^{4}p}{p^{2}-m^{2}_{q}} \ + \
3g'\int\frac{d^{4}p}{p^{2}-m^{2}_{q}} \ = \ 0 .
\end{equation}
\begin{figure}
\centering
\includegraphics[width=8cm]{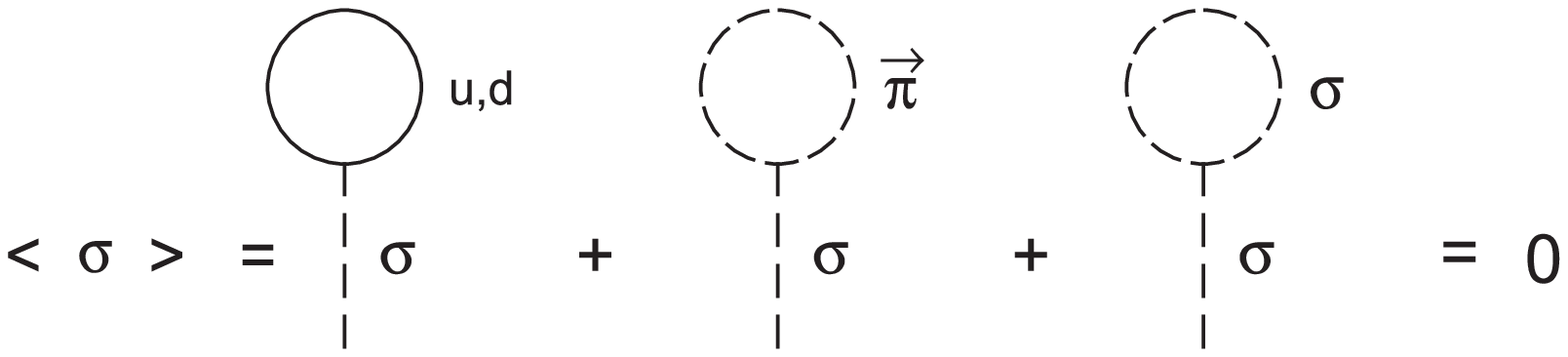}
\caption{\textit{Ben~Lee null tadpole sum}.}
\end{figure} 
Here we have dropped the small pion loop (due to the Nambu-Goldstone
theorem), used $8=4 \times 2$ with 4 from the fermion trace, 2 from
u,d flavours, and invoked the combinatoric factor of 3 for the $\sigma
- \sigma -\sigma$ coupling.
The false vacuum\cite{r1} has \mbox{$<\sigma> = f_{\pi} \ \ne \ 0,\ \
<\pi> = 0$} , whereas after the shift the true vacuum $<\sigma> =
<\pi> = 0$ is characterized by Fig.~1 and by Eq.~(\ref{e22}).  We
solve Ben~Lee's equation (\ref{e22}) above by using the L$\sigma$M
chiral relations $g = m_{q}/f_{\pi}$ and $g' =
m^{2}_{\sigma}/2f_{\pi}$ from Eq.~(\ref{e12}) together with simple
dimensional analysis requiring quadratic divergence mass scales
\begin{equation} \label{e23}
\int \frac{d^{4}p}{p^{2}-m^{2}_{q}} \sim m^{2}_{q}
, \hspace{0.25in}
\int \frac{d^{4}p}{p^{2}-m^{2}_{\sigma}} \sim m^{2}_{\sigma}.
\end{equation}
Thus Eq.~(\ref{e22}) above implies\cite{r3}
%
%
\begin{equation} \label{e24}
N_{c}(2m_{q})^{4} \ = \ 3m^{4}_{\sigma},
\end{equation}
so the NJL value $m_{\sigma}= 2m_{q}$ requires $N_{c}= 3$ in
Eq.~(\ref{e24}).  The SU(2) L$\sigma$M also requires the loop value
$m_{\sigma} = 2m_{q}$, which we derive in the next section, and again
$N_{c}=3$ follows from the L$\sigma$M null tadpole condition
Eq.~(\ref{e22}) leading to Eq.~(\ref{e24}).

\section{Log-Divergent Gap Equation (LDGE) in the L$\sigma$M}
\begin{figure}
\centering
\includegraphics[width=8cm]{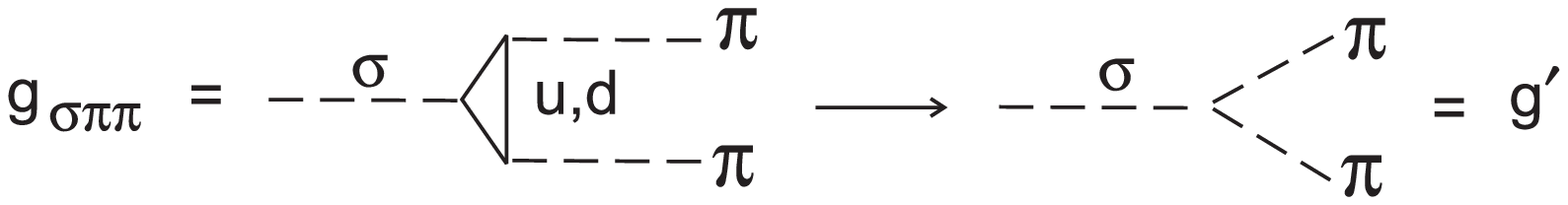}
\caption{\textit{L$\sigma\!$M cubic meson log-divergent coupling
$g_{\sigma \pi \pi} \rightarrow g'$}.}
\end{figure} 

Defining the pion decay constant via $< 0 \ | A_{\mu}^{i}|\pi^{j} > =
if_{\pi}q_{\mu}\delta^{ij}$, the u,d quark loops combined with the GTR
$f_{\pi}g = m_{q}$ leads to the LDGE\cite{r10} in the CL :
\begin{equation} \label{e31}
-i4N_{c}g^{2}\int \frac{ \overline{d}^{4}p }{(p^{2}-m^{2}_{q})^{2} } \
= \ 1.
\end{equation}
It is a gap equation because the ``mass gap'' $m_{q}$ cancels out to
form (\ref{e31}).  This LDGE Eq.~(\ref{e31}) also follows from the NJL
analysis.  Morever, the u,d quark loops for the pion form factor
(using $g\gamma_{5}$ pseudoscalar scaling) in turn requires \linebreak
$F_{\pi}(q^{2}=0)=1$ since then PVV quark loops predict the pion form
factor\cite{r11}
\begin{equation} \label{e32}
F_{\pi}(q^{2} ) \ =\ -i4N_{c}g^{2} \int_{0}^{1} \ dx \ \int \frac{
\overline{d}^{4}p } {[ p^{2} - m^{2}_{q} + x(1-x)q^{2}]^{2} }\ \ ,
\end{equation}
for $\overline{d}^{4}p = d^{4} p/(2\pi)^{4}$.  Clearly Eq.~(\ref{e32})
reduces to the LDGE Eq.~(\ref{e31}) when \mbox{$F_{\pi}(q^{2}=0)=1$}.

Next, the 3-point function $g_{\sigma \pi \pi}$ quark triangle in
Fig.~2 reduces in the CL to\cite{r3}
%
%
%
%
%
%
\begin{equation} \label{e33}
g_{\sigma \pi\pi} \ = \ - 8ig^{3}N_{c}m_{q} \int \frac{ \overline{d}^{4}p
} {[ p^{2} - m^{2}_{q} ]^{2} } \ = \ 2gm_{q} \ \ ,
\end{equation}
by virtue of the LDGE (\ref{e31}) and then the GTR, $m_{q} = f_{\pi}g$
``shrinks'' (\ref{e33}) to
\begin{equation} \label{e34}
g_{\sigma \pi\pi} \ = \ 2m_{q} g \ = \ (2m_{q})^{2}/ 2f^{CL}_{\pi} \ = \
m^{2}_{\sigma}/2f_{\pi} \ = \ g',
\end{equation}
the tree-level L$\sigma$M value in Eq.~(\ref{e12}).  This is valid
\emph{only if} $m_{\sigma}= 2 m_{q}$ for the loop-level
L$\sigma$M~\cite{r3}.

Also, the 4-point function $g_{\pi \pi \pi \pi}$ quark box in Fig.~3
shrinks in the CL to\cite{r3}
\begin{figure}
\centering
\includegraphics[width=8cm]{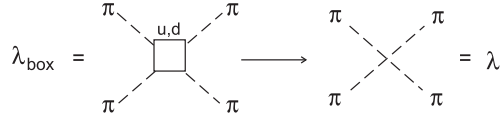}
\caption{\textit{L$\sigma \!$M quartic meson log-divergent coupling
$\lambda_{box} \rightarrow \lambda $}.}
\end{figure} 
\begin{equation} \label{e35}
\lambda_{box} \ = \ -8iN_{c}g^{4} \int \frac{\overline{d}^{4}p }
{[ p^{2} - m^{2}_{q} ]^{2} } \ =\ 2 g^{2} \ = \ g'/f_{\pi} \ =
\ \lambda \ ,
\end{equation}
the L$\sigma$M contact term in Eq.~(\ref{e12}), but again we must
invoke the LDGE (\ref{e31}) and $m_{\sigma} = 2m_{q}$.  In effect, the
quark loops of Figs.~1,2,3 dynamically reproduce the entire SU(2)
L$\sigma$M Lagrangian with $N_c=3$ and $g = 2\pi/\sqrt{3}$, $g^\prime
= 2g \, m_q$, and $\lambda = 2g^2 = 8\pi^2/3$.  Note that the large
$N_{c}$ limit for a $U_{L}(1) \times U_{R}(1)$ model
predicts\cite{r12} $m_{\sigma}
\rightarrow 2m_{q}$ but $\lambda \rightarrow g^{2}$ (not $2g^{2}$) as
$N_{c} \rightarrow \infty$.
%
\section{Chiral phase transition}
An independent test of the above SU(2) L$\sigma$M stems from the
chiral phase transition approach, where the thermalized fermion
(quark) propagator is replaced at finite temperature by\cite{r13}
\begin{equation} \label{e41}
\frac{ i(\gamma \cdot p + m) }{p^{2} - m^{2}  } \ \rightarrow\
\frac{ i(\gamma \cdot p + m) }{p^{2} - m^{2} }\  -\ \frac{ 2\pi
\delta (p^{2} - m^{2})(\gamma \cdot p + m) }{e^{ |P_{0}/ T| }+1  }
\ \ . 
\end{equation}
This approach leads to the chiral restoration (melting) temperature
$T_{c}$, which corresponds to a second-order phase transition.  Over
the past 15 years, many authors have predicted for two quark flavours
that this order parameter is\cite{r14}
\begin{equation} \label{e42}
T_{c} \ = \ 2f^{CL}_{\pi} \ = \ 2 \times 90 \ {\rm MeV} \ = \ 180 \
{\rm MeV} \ \ ,
\end{equation}
where the 3\% reduction of $f_{\pi}$ in Eq.~(\ref{e13}) above has been
used.  Only recently, F. Karsch\cite{r15} has ``measured'' $T_{c}$ via
the computer lattice to find for $N_{f}=2$:
\begin{equation} \label{e43}
T_{c} \ = \ 173 \ \pm \ 8 \ {\rm MeV} \ \ ,
\end{equation}
so the various melting studies theoretically predicting the nearby
Eq.~(\ref{e42}) should be reviewed in detail.

\subsection{Melting the SU(2) Quark Mass}

As $T$ increases to the chiral ``melting'' temperature $T_{c}$, the
quark mass melts, and the $\sigma$ tadpole graph of Fig.~4 via
Eq.~(\ref{e41}) gives\cite{r16}
\begin{figure}
\centering
%
\includegraphics[width=4cm]{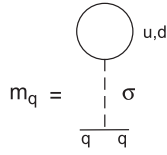}
\vspace{-1.0cm}
\caption{\textit{L$\sigma$M quark tadpole representation of quark mass
$m_{q}$}.}
\end{figure} 
\begin{equation} \label{e44}
0 \ \leftarrow \ m_{q}(T_{c}) \ = \ m_{q} \ + \ \frac{
8N_{c}g^{2}m_{q} }{-m_{\sigma}^{2} } \left( \frac{ T_{c}^{2}
}{2\pi^{2} } \right) J_{+} \ \ ,
\end{equation}
where we use $N_{c}=3$ and the integral $J_{+} = \int_{0}^{\infty} x
dx [e^{x}+1]^{-1}=\pi^{2}/12.$  Then Eq.~(\ref{e44}) reduces to
\begin{equation} \label{e45}
T_{c}^{2} \ = \ m^{2}_{\sigma}/g^{2} \ \ .
\end{equation}
Note that the NJL-L$\sigma$M relation $m_{\sigma}=2m_{q}$ combined
with the GTR \mbox{$g=m_{q}/f_{\pi}^{CL}$} means that the square root
of (\ref{e45}) is $T_{c} = 2f_{\pi}^{CL}$, Eq.~(\ref{e42}) above.

\subsection{Melting the Scalar $\sigma$ Mass in the L$\sigma $M}

Assuming the Lagrangian density quartic coupling is proportional to
\mbox{${\cal L} \sim -(\lambda/4)\sigma^{4},$} but with
$J_{+} \rightarrow J_{-} = \int_{0}^{\infty} x dx [e^{x}-1]^{-1}=
\pi^{2}/6,$ Refs.~[14] then find
\begin{equation} \label{e46}
T_{c}^{2} \ = \ 2m^{2}_{\sigma} / \lambda \ \ ,
\end{equation}
in an SU(2) L$\sigma$M context.  Then, independent of the value of
$T_{c}$ or $m_{\sigma}$, Eqs.~(\ref{e45}), (\ref{e46}) together
require $\lambda=2g^{2}$, as already found in Eq.~(\ref{e35}) in
L$\sigma$M loop order.

\subsection{Melting the Quark Condensate in QCD}

Noting that thermalizing the quark condensate as
$\delta<\overline{q}q> = <\overline{q}q>\delta \overline{q}
\ + \linebreak <\overline{q}q>\delta q$, Ref.~[17] finds in QCD
\begin{equation} \label{e47}
T_{c}^{2} \ = \ 3 m^{2}_{q}/ \pi^{2} \ \ .
\end{equation}
Expressing the square root of Eq.~(\ref{e47}) together with
Eq.~(\ref{e45}) and again using the NJL-L$\sigma$M relation
$m_{\sigma} = 2 m_{q}$ leads to the on-shell meson-quark coupling
(with $N_{c}=3$)
\begin{equation} \label{e48}
g \ = \ 2\pi/\sqrt{3}, \ \ 
\end{equation}
as already predicted in Sec.~1 for the L$\sigma$M.

In QCD language, $\alpha_{s}(1\ {\rm GeV}) \approx 0.5$\cite{r18}
grows at $m_{\sigma} \sim$ 600 MeV to $\alpha_{s}(m_{\sigma})|_{\rm
QCD} \approx \pi/4$, the infrared limit\cite{r19}.  Then the $C_{2F}$
factor increases $\alpha_{s}$ to $\alpha_{\rm QCD}^{\rm
eff}(m_{\sigma}) = (4/3)\pi/4 = \pi/3$, i.e., to quark condensation
with $\alpha_{s}^{\rm eff} \approx 1$.  In L$\sigma$M language
Eq.~(\ref{e48}) corresponds to $g^{2}/4\pi = \pi/3$, the above
infrared value of QCD\cite{r20}.  Note that the latter has already
been derived in Eq.~(\ref{e14}) via the loop-order
L$\sigma$M~\cite{r3} or via the Z = 0 compositeness
condition\cite{r5}.

\section{Invoking the Dim.\ Reg.\ Lemma for Loop-Order L$\sigma$M}
Lastly, we can self-consistently nonperturbatively derive the
NJL-L$\sigma$M relation $m_{\sigma} = 2m_{q}$.  The constituent quark
mass $\sigma$ tadpole graph of Fig.~4 generates the quadratically
divergent mass gap equation (QDGE)
\begin{equation} \label{e51}
m_{q} \ = \ - \frac{ 8iN_{c}g^{2} }{m^{2}_{\sigma} } \int \frac{
\overline{d}^{4}p \ m_{q}}{p^{2}\ -\ m^{2}_{q} } \ \ .
\end{equation}
Such a quadratic divergence also appears in the dimensional
regularization (dim.\ reg.) lemma (DRL)
\begin{equation} \label{e52}
\int \overline{d}^{4}p \left[ \frac{ m_{q}^{2} }{ (p^{2} - m^{2}_{q})^{2} }
- \frac{ 1}{p^{2} - m^{2}_{q} } \right] = \lim_{ l \rightarrow 2}
\frac{ im_{q}^{2l - 2} }{( 4 \pi)^{2} }[ \Gamma(2-l)+\Gamma ( 1-l) ]=-
\frac{i m_{q}^{2}}{ (4 \pi)^{2} },
\end{equation}
due to the gamma function identity\cite{r3} $\Gamma(2-l) +
\Gamma(1-l) \rightarrow -1$ as $2l \rightarrow 4$.  Combining the DRL
Eq.~(\ref{e52}) with the QDGE Eq.~(\ref{e51}) and the LDGE
Eq.~(\ref{e31}) and dividing out the nonzero quark mass (gap), one
obtains\cite{r3}
\begin{equation} \label{e53}
1 \ = \ \frac{ 2m_{q}^{2} }{m_{\sigma}^{2} } \left[ 1 \ + \
\frac{g^{2}N_{c}}{4 \pi^{2}} \right] \ \ .
\end{equation}

Further, solving Eq.~(\ref{e53}) as 1 = 1/2 + 1/2 (due to the pion and
$\sigma$ bubbles giving $m_{\pi}^{2} = 0, \ m_{\sigma}^{2} =
N_{c}g^{2}m^{2}/\pi^{2})$, one finds\cite{r3} the NJL-L$\sigma$M and Z
= 0 loop-order solutions of Eqs.~(\ref{e14}).  The above dim.\ reg.\
formalism is

$a$) compatible with the LDGE ``nonperturbative shrinkage'' of the quark
triangle and box L$\sigma$M Eqs.~(\ref{e34}) and (\ref{e35}) for any
value of $N_{c}$, giving again $m_{\sigma} = 2 m_{q}$, $g = 2\pi/
\sqrt{N_{c}}$,

$b$) compatible with the chiral phase transition approach of Sec.~4,
but only when $N_{c}=3$,

$c$) compatible with analytic, zeta function, and Pauli-Villars
regularization schemes when the massless quadratic loop analogue of
Eqs.~(\ref{e23}) holds\cite{r21}:
\begin{equation} \label{e54}
\int d^{4}p / p^{2} \ = \ 0 \ \ ,
\end{equation}
again due to simple dimensional analysis.

Recall that the massive SU(2) loop versions Eqs.~(\ref{e23}) allowed
us to theoretically deduce that $N_{c} = 3$ due to the B.W. Lee null
tadpole sum Eq.~(\ref{e22}), leading to Eq.~(\ref{e24}) because we
already know $m_{\sigma} = 2m_{q}$ in the NJL or L$\sigma$M theories
at loop order.

\section{Conclusion}

Prof.\ 't Hooft began his film lectures by first writing the
L$\sigma$M Lagrangian, but instead let $N_{c} \rightarrow \infty$ in
order to perturbatively solve the L$\sigma$M and also QCD\cite{r22}.
While we have no objection to such an approach, we suggest it in fact
is possible to start with the usual value $N_{c} = 3$ and instead
nonperturbatively solve in loop order the SU(2) L$\sigma$M with the
NJL--Ben~Lee null tadpole solution $m_{\sigma} = 2m_{q}$.  This also
leads to L$\sigma$M nonperturbative shrinkage in Sec.~3 via the LDGE,
to the chiral phase transition temperature $T_{c} = 2 f_\pi^{CL}
\approx 180$ MeV for $N_{f}=2$, and to the formal LDGE-DRL-QDGE
solution $m_{\sigma} = 2m_{q},\ g = 2\pi/\sqrt{N_{c}}$ in Sec.~4.  Of
course the L$\sigma$M requires $g^{2}N_{c}$ = constant $(4\pi^{2})$
with amplitudes scaling like $1/N^{2}_{c} \ (1/9)$, so 't Hooft's
large $N_{c}$ result that $1/N_{c}^{2}$ terms vanish as $N_{c}
\rightarrow \infty$ is 90\% true anyway when $N_{c}=3$.

\subsection*{Acknowledgements}
The author thanks R. Delbourgo, J. Cleymans, V. Elias and
P. Zenczykowski for prior discussions.

\end{document}